\documentclass[aps,prc,showpacs,superscriptaddress,groupedaddress,twocolumn]{revtex4}
\usepackage{feynmp}
\usepackage{amsmath}
\usepackage{amssymb}
\usepackage{graphicx}

\newcommand{\sci}[2]{#1\times 10^{#2}}
\newcommand{\scit}[2]{#1 \, \left(#2\right)}

\newcommand{\be}{\begin{equation}}
\newcommand{\ee}{\end{equation}}
\newcommand{\bea}{\begin{eqnarray}}
\newcommand{\eea}{\end{eqnarray}}
\newcommand{\ba}{\begin{array}}
\newcommand{\ea}{\end{array}}

\long\def\symbolfootnote[#1]#2{\begingroup%
\def\thefootnote{\fnsymbol{footnote}}\footnote[#1]{#2}\endgroup}

\def\apj{ApJ }
\def\apjs{ApJ  Supplement}

\def\apjs{ApJS }

\def\aap{A\&A }

\usepackage{hyperref}

\begin{document} 
{\begin{flushleft} INT-PUB-12-024
\vskip -0.5in
{\normalsize }
\end{flushleft}
\title{Medium modification of the charged current neutrino opacity and its implications}
\author{L. F. Roberts}
\affiliation{Department of Astronomy and Astrophysics, University of 
California, Santa Cruz, California 95064, USA}
\email{lroberts@ucolick.org}
\author{Sanjay Reddy}
\affiliation{Institute for Nuclear Theory, University of Washington, 
Seattle, Washington 98195, USA}
\email{sareddy@uw.edu}
\author{Gang Shen}
\affiliation{Institute for Nuclear Theory, University of Washington, 
Seattle, Washington 98195, USA}
\email{wolfgang.shen@gmail.com}

\begin{abstract}

Previous work on neutrino emission from proto-neutron stars which employed
full solutions of the Boltzmann equation showed that the average energies 
of emitted electron neutrinos and antineutrinos are closer to one another than 
predicted by older, more approximate
work.  This in turn implied that the neutrino driven wind is proton rich during 
its entire life, precluding $r$-process nucleosynthesis and the synthesis of Sr, Y, and Zr.  
This work relied on charged current neutrino interaction rates that are appropriate 
for a free nucleon gas.  Here, it is shown in detail that the 
inclusion of the nucleon potential energies and collisional broadening of the response  
significantly alters this conclusion.  Iso-vector interactions, which give rise to the 
nuclear symmetry energy, produce a difference between the neutron and proton single-particle 
energies $\Delta U=U_n-U_p$ and alter the kinematics of the charged current reactions. In 
neutron-rich matter, and for a given neutrino/antineutrino energy, the rate for 
$\nu_e+n\rightarrow e^-+p$ is enhanced while  $ \bar{\nu}_e+p\rightarrow n+e^+$ is 
suppressed because the $Q$ value for these reactions is altered by $\pm\Delta U$, respectively.  
In the neutrino decoupling region, collisional broadening acts to enhance both $\nu_e$ and 
$\bar{\nu}_e$ cross-sections and RPA corrections decrease the $\nu_e$ cross-section and increase 
the $\bar \nu_e$ cross-section, but mean field shifts have a larger effect.  Therefore, electron 
neutrinos decouple at lower temperature than when the nucleons are assumed to be free and have 
lower average energies.  The change is large enough to allow for a reasonable period of time 
when the neutrino driven wind is predicted to be neutron rich.  It is also shown that the 
electron fraction in the wind is influenced by the nuclear symmetry energy.
\end{abstract}

\maketitle 

\section{Introduction} 

The neutrino opacity of dense matter encountered in core-collapse supernova is of 
paramount importance to the explosion mechanism, potential nucleosynthesis, 
supernova neutrino detection and to the evolution of the compact remnant left behind.  
Matter degeneracy, strong and electromagnetic 
correlations, and multi-particle excitations have all been shown to be important, 
especially at supra-nuclear densities \citep[e.g.][]{Reddy98,Burrows98,Burrows99,
Reddy99,Hannestad98,Horowitz03,Lykasov08,Bacca11}.  Supernova and proto-neutron star (PNS) 
simulations that employ some subset of these improvements to the free gas neutrino interaction 
rates have found that these corrections play a role in shaping the temporal and spectral 
aspects of neutrino emission \citep{Pons99,Reddy99,Huedepohl10,Roberts12}.  Much is still 
uncertain, especially because of the approximations one must make regarding weak interactions
with the dense background medium.  A specific issue 
of importance is the difference between the average energies of electron neutrinos and electron 
antineutrinos. This difference is largely determined by the charged current reactions 
$\nu_e+n\rightarrow p+ e^-$ and  $\bar{\nu}_e+p\rightarrow n + e^+ $  in neutron-rich matter 
at densities $\rho \simeq 10^{12}-10^{14}$ g/cm$^3$.

Recently, one of the authors has shown that an accurate treatment of mean field effects in 
simulations of PNS cooling changes the predicted electron fraction in the 
neutrino driven wind (NDW) \citep{Roberts12b} relative to simulations which do not account 
for mean field potentials in nuclear matter \citep{Fischer10,Huedepohl10,Fischer12}.
This difference has significant consequences for the nucleosynthesis expected in the NDW 
\citep[e.g.][]{Hoffman97,Roberts10,Arcones11} and for neutrino oscillations outside the neutrino
sphere \cite{Duan06,Duan10}. In this work, we discuss generic aspects of strong 
interactions that lead to a large asymmetry in the charged current reaction rates for electron neutrinos
and antineutrinos. We also demonstrate that this difference manifests itself in potentially 
observable effects on neutrino spectra from supernovae and that the difference depends on the 
assumed density dependence of the nuclear symmetry energy.  The effect of multi-particle 
excitations and correlations (via the RPA) on the charged current response are also explored.

Neutron-rich matter at densities and temperatures relevant to the neutrino sphere of a PNS is characterized 
by degenerate relativistic electrons and non-relativistic partially degenerate neutrons and 
protons.  Beta-equilibrium, with net electron neutrino number $Y_{\nu_e}=0$ is a reasonably 
good approximation for the material near the neutrino sphere because, by definition, this 
material can efficiently lose net electron neutrino number.  At these densities, effects due 
to strong interactions modify the equation of state and the beta-equilibrium abundances of 
neutron and protons. Simple models for the nuclear equation of state predict that the nucleon 
potential energy is
\be 
U_{n/p} \approx V_{\text{is}}~(n_n+n_p)\pm  V_{\text{iv}}~(n_n-n_p)\,, 
\ee
where $V_{\text{is}}$ and $V_{\text{iv}}$ are the effective iso-scalar and iso-vector potentials. 
Empirical properties of nuclear matter and neutron-rich matter suggest that $V_{\text{is}}\times 
n_0\approx -50$ MeV and $V_{\text{iv}}\times n_0\approx 20$ MeV.  The potential energy associated 
with $n \rightarrow p$ conversion in the medium is
\be 
\Delta U = U_n-U_p \approx 40 \times \frac{(n_n-n_p)}{n_0}~ {\rm MeV}, 
\ee
where $n_0=0.16$ nucleons/fm$^3$ is the number density at saturation.  It will be shown that $\Delta U$ 
changes the kinematics of charged current reactions, so that the $Q$-value for the reaction $\nu_e+n\rightarrow e^-+p$ 
is enhanced by $\Delta U$ while that for $\bar{\nu}_e+p\rightarrow e^+ + n$ is reduced by the same 
amount.  The effect is similar to the enhancement due to the neutron-proton mass difference, but 
is larger when the number density $n > n_0/20$. 

In section \ref{sec:CCresponse}, charged current neutrino opacities in an interacting medium are 
discussed.  We consider how mean fields affect the response of the medium in detail and how this 
depends on the properties of the nuclear equation of state.  The effects of nuclear correlations 
and multi-particle hole excitations are also discussed.  In section \ref{sec:PNS_evolve}, the 
effect of variations of the charged current reaction rates on the properties of the emitted 
neutrinos is studied.

\section{The Charged Current Response}
\label{sec:CCresponse}
The differential absorption rate for electron neutrinos by the process 
$\nu_e+n\rightarrow e^-+p$ is given by 
\bea 
\frac{1}{V}
\frac{d^2{\sigma}}{d\cos\theta dE_e} = \frac{G_F^2 \cos^2 \theta_c}{4 \pi^2} p_e~E_e \, (1-f_e(E_e))  \nonumber\\
\times \left[(1 + \cos\theta)   S_\tau(q_0,q)+ g_A^2(3-\cos\theta) S_{\sigma \tau}(q_0,q)\right]\,
\label{eq:dcross}
\eea
where $S_\tau(q_0,q)$ and $S_{\sigma \tau}(q_0,q)$ are the response functions associated with the Fermi and Gamow-Teller operators, $\tau_+$ and $\sigma \tau_+$, respectively.  The energy transfer to the nuclear medium is $q_0 = E_\nu-E_e$, and 
the magnitude of the momentum transfer to the medium is $q^2 = E_\nu^2 + 
E_e^2 - 2E_\nu E_e\cos\theta$.  In a non-interacting Fermi gas, the response functions $S_\tau(q_0,q) =S_{\sigma \tau}(q_0,q)
=S_{\rm F}(q_0,q)$ are given by 
\bea
S_{\rm F}(q_0,q) = \frac{1}{2\pi^2} \int d^3p_2 \delta(q_0 + E_2 - E_4)f_2(1-f_4),
\label{eq:response}
\eea
where the particle labeled 2 is the incoming nucleon and the particle labeled 4 
is the outgoing nucleon.  When the dispersion relation for nucleons is given by 
$E(p)=M+p^2/2M$ -- neglecting the neutron-proton mass difference for simplicity -- 
the integrals in Eq.~\ref{eq:response} can be performed to obtain 
\be
S_{\rm F}(q_0,q) =  \frac{2}{1 - e^{-z}}~\text{Im}~\Pi_\text{F}
\ee
where $z= (q_0 + \mu_2- \mu_4)/T$ and 
\be 
\text{Im}~\Pi_\text{F} = \frac{M^2 T}{2 \pi q} \ln\left\{
\frac{\exp\left[\left(e_{\rm min} - \mu_2\right)/T  \right] + 1}
{\exp\left[\left(e_{\rm min} - \mu_2\right)/T \right]+\exp\left[-z\right]} 
\right\}\,,
\label{eq:pi_fg}
\ee
is the free particle-hole polarization function.  $\mu_2$ and $\mu_4$ are the chemical 
potentials of the incoming and outgoing nucleons, $M$ is the nucleon mass,   
and
\be
e_{\rm min} = \frac{M}{2q^2 }\left(q_0-\frac{q^2}{2M}\right)^2\,.
\label{eq:emin}
\ee
$e_{\rm min}$ arises from the kinematic restrictions imposed by
energy-momentum transfer and the energy conserving delta function. 
Physically, $e_{\rm min}$ is the minimum energy of the nucleon in the initial 
state that can accept momentum $q$ and energy $q_0$.    

\subsection{Frustrated Kinematics}
\label{sec:kinematics}
The differential cross-section of $\nu_e$ absorption is the product of the nucleon 
response times the available electron phase space
\be 
p_e~E_e~(1-f_e(E_e))  \approx E_e^2~\exp{\left(\frac{E_e-\mu_e}{T}\right)}  \,. 
\label{eq:phase_space}
\ee  
Due to the high electron degeneracy, the lepton phase space increases exponentially 
with the electron energy.  To completely overcome electron blocking requires $E_e=E_{\nu_e}-
q_0\approx \mu_e$ or $q_0\approx -\mu_e$ when $E_{\nu_e} \ll \mu_e$. However, the fermi gas response 
function in Eq.~\ref{eq:response} is peaked at $q_0\simeq q^2/2M   \approx 0$ reflecting 
the fact that nucleons are heavy.  At large $|q_0|\simeq q \approx \mu_e$ the response 
is exponentially suppressed due to kinematic restrictions imposed by Eq. ~\ref{eq:emin} 
which implies only neutrons with energy 
\be 
E_2  > e_{\rm min}  \simeq  \frac{M}{2q^2}~q^2_0 \approx \frac{M}{2}\,,
\ee
can participate in the reaction. For conditions in the PNS decoupling region, and in 
the fermi gas approximation, the $\nu_e$ reaction proceeds at  $q_0\approx 0$ at the 
expense of large electron blocking.  Thus effects that can shift strength to more 
negative $q_0$ can increase the electron absorption rate exponentially.   

It is well known that the neutron-proton mass difference $\Delta M=M_n-M_p$ increases 
the $Q$ value for this reaction and a more general expression for $S(q_0,q)$ derived 
in \cite{Reddy98} includes this effect. The effect of $\Delta M$ can be understood by 
noting, that at leading order, it only changes the argument of the energy delta-function 
in Eq. \ref{eq:response} and is subsumed by the replacements $q_0 
\rightarrow (q_0+ \Delta M)$ and 
\be 
e_{\rm min} \rightarrow \tilde{e}_{\rm min}\approx \frac{M}{2q^2 }
\left(q_0+\Delta M -\frac{q^2}{2M}\right)^2\,. 
\ee
This shift changes the location of the peak of the response by moving it to the 
region where $E_e$ is larger and confirming that it increases 
the $Q$ value and the final state electron energy by $\Delta M=M_n-M_p$. From 
Eq.~\ref{eq:phase_space} we see that the rate for $\nu_e$ absorption is increased by 
roughly a factor $(1+\Delta M/E_e)^2~\exp{(\Delta M/T)}$. By the same token, the $Q$ 
value for the reaction $\bar{\nu}_e+p\rightarrow e^++n$ is reduced by $\Delta M$ and 
this acts to reduce the rate. In this case, the detailed balance factor $[1-\exp{(-z)}]^{-1}$ 
in the response function $S(q_0,q)$ is the source of exponential suppression -- simply 
indicating a paucity of high energy protons in the plasma. For small $q_0\ll \mu_e$, the 
detailed balance factor is
\be  
\frac{-1}{1-\exp{(-z)}}\approx \exp{\left(\frac{q_0-\mu_e}{T}\right)},
\ee
where we have used the fact that $\mu_n-\mu_p=\mu_e$ in beta-equilibrium. Since $q_0 
\rightarrow (q_0-\Delta M)$ for the $\bar{\nu}_e$ process,  $\Delta M$ will suppress 
this rate exponentially. This is in line with the expectation that $\Delta M$ increases the 
cross-section for $\nu_e$ absorption and decreases it for $\bar{\nu}_e$ absorption. 
In the following we show that the mean field energy shift, driven by the nuclear symmetry 
energy, has a similar but substantially larger effect in neutron-rich matter at densities 
$\rho \gtrsim 10^{12}$ g/cm$^3$. 
 
\subsection{Mean Field Effects}

\begin{figure}
\begin{center}
\includegraphics[width=\columnwidth]{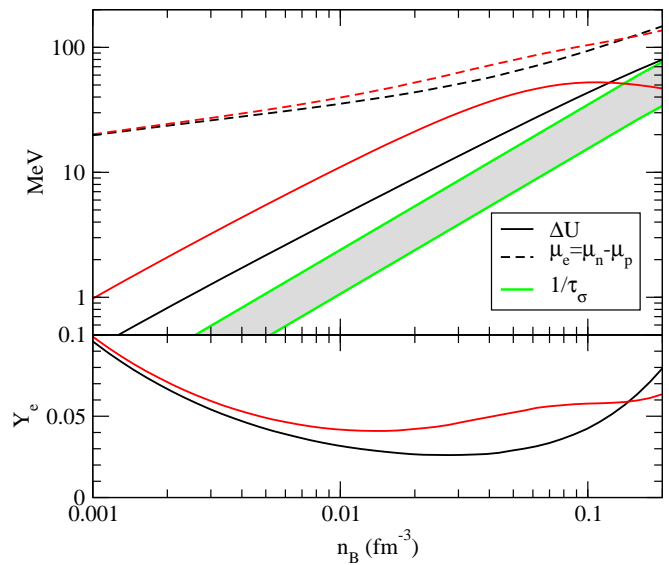}
\caption{ {\it Top Panel:} The electron chemical potential (dashed lines) and $\Delta U=U_n-U_p$  
(solid lines) are shown as a function of density for the two equation of state models (IUFSU: red curves and GM3: black curves) in beta-equilibrium for $Y_\nu = 0$  and $T=8$ 
MeV. The grey band shows an approximate range of values for inverse spin relaxation time calculated 
in \cite{Bacca11} and is discussed in connection with collisional broadening. {\it Bottom Panel:} 
The equilibrium electron fraction as a function of density for the two equations of state shown 
in the top panel.}
\label{fig:potentials}
\end{center}
\end{figure} 

Interactions in the medium alter the single particle energies, and nuclear mean field 
theories predict a nucleon dispersion relation of the form 
\be
E_i(k) = \sqrt{k^2 + M^{*2}} + U_i \equiv K(k) + U_i\,, 
\label{eq:mf_dispersion}
\ee
where $M^*$ is the nucleon effective mass and $U_i$ is the mean field energy shift. For neutron-rich 
conditions, the neutron potential energy is larger due to the iso-vector nature of the strong interactions. 
The difference $\Delta U=U_n-U_p$ is directly related to the nuclear symmetry energy, which is the difference between the energy 
per nucleon in neutron matter and symmetric nuclear matter.  Ab-intio methods using Quantum Monte Carlo reported in 
\cite{Akmal1998} and \cite{Gandolfi12}, and chiral effective theory calculations of neutron matter by \cite{Hebeler10} suggest 
that the symmetry energy at sub-nuclear density is larger than predicted by many mean field models currently employed in supernova 
and neutron star studies (for a review see \cite{Steiner05}). To highlight the symmetry energies importance, we choose two models for the dense matter equation of state: (i) the 
GM3 relativistic mean field theory parameter set without hyperons \citep{Glendenning91} where the symmetry energy is linear at low 
density; and (ii) the IU-FSU parameter set \citep{Fattoyev10} where the symmetry energy is non-linear in the density and large at sub-nuclear 
density.  

The electron chemical potential (dashed lines) and neutron-proton potential energy difference (solid lines) for these two models 
are shown as a function of density in beta-equilibrium in Figure \ref{fig:potentials}. Here $Y_\nu = 0$ for all densities and a temperature of 8 MeV is assumed.  At sub-nuclear densities, the IU-FSU $\Delta U$ is always larger than the GM3 $\Delta U$ 
value due to the larger sub-nuclear density symmetry energy in the former.  The electron chemical potential as a function of density, 
as well as the equilibrium electron fraction, is shown in Figure \ref{fig:potentials} for both models.  In beta-equilibrium, models 
with a larger symmetry energy predict a larger electron fraction for a given temperature and density.  Therefore, IU-FSU has a larger equilibrium $\mu_e$ than GM3 and the reaction $\nu_e + n \rightarrow e^- + p$ will experience 
relatively more final state blocking. However, as we show below, the inclusion of $\Delta U$ in the reaction kinematics is needed for consistency. 

To elucidate the effects of $\Delta U$ we set $M^*=M$ and note that this assumption can easily be relaxed 
\citep{Reddy98} and it does not change the qualitative discussion below.  Because in current equation of state models the potential, $U_i$, is independent of the momentum, $k$, this form of the dispersion relation results in a free Fermi gas distribution function with single particle energies $K(k)$ for 
nucleons of species $i$, but with an effective chemical potential $\tilde \mu_i \equiv \mu_i - U_i$.  This fact was emphasized in 
\citep{Burrows98}, and used to show that it was unnecessary to explicitly know the values of the nucleon potentials for a given nuclear 
equation of state (which are often not easily available from widely used nuclear equations of state in the core-collapse supernova community) 
when calculating the neutral current response of the nuclear medium.  Clearly, if both $\mu_i$ and $\tilde \mu_i$ are known, then $U_i$ can be 
easily obtained.  This implies that for a given temperature, density and electron fraction, the neutral current response function is unchanged 
in the presence of mean field effects, as the kinematics of the reaction are unaffected by a constant offset in the nucleon single particle energies.  
In contrast, the kinematics of the charged current reaction are affected by the difference between the neutron and proton potentials and the charged 
current response is altered in the presence of mean field effects. 

Inspecting the response function in Eq.~\ref{eq:response} and the dispersion relation in Eq.~\ref{eq:mf_dispersion} 
it is easily seen that the mean field response is
\be
S_{\rm MF} (q_0,q) = \frac{2}{1 - e^{-z}}~\text{Im}~\Pi_\text{MF}
\ee
where 
\be 
\text{Im}~\Pi_\text{MF}=\frac{M^2 T}{2 \pi q} \ln\left\{
\frac{\exp\left[\left(\tilde e_{\rm min} - \tilde \mu_2\right)/T  \right] + 1}
{\exp\left[\left(\tilde e_{\rm min} - \tilde \mu_2\right)/T \right]+\exp\left[-z\right]} 
\right\}\,,~~~
\label{eq:mf_response}
\ee
and 
\be
\tilde e_{\rm min} = \frac{M}{2q^2}(q_0+U_2-U_4-q^2/2M)^2
\label{eq:mf_emin}
\ee
This is obtained from the free gas response 
by the replacements 
\bea
\label{eq:mf_replacements}
\mu_i &\rightarrow& \tilde \mu_i = \mu_i - U_i\nonumber \\
q_0 &\rightarrow& \tilde q_0 = q_0 + U_2 - U_4 
\eea 
and $q \rightarrow q$.   
Therefore, we see that the potential difference $\Delta U= \pm(U_2 - U_4)$ affects reaction kinematics and cannot be subsumed 
in the redefinition of the chemical potentials (to yield the same individual number densities). 

\begin{figure}
\begin{center}
\leavevmode
\includegraphics[width=\columnwidth]{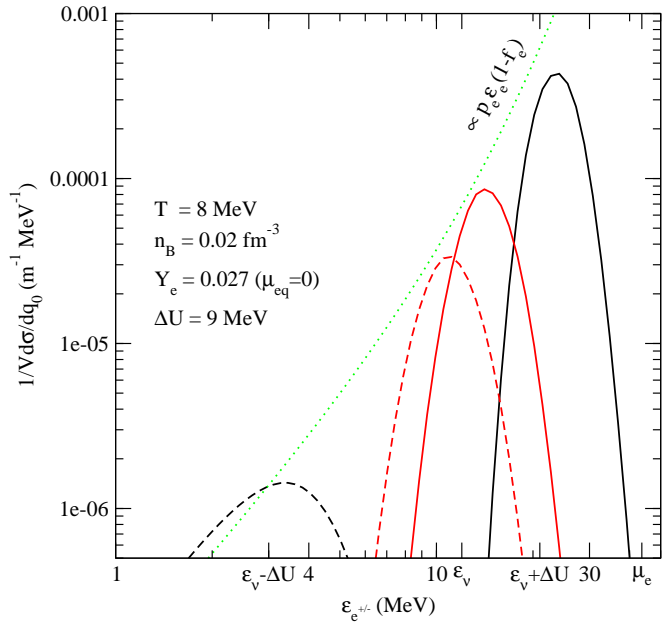}
\caption{ Angle integrated differential cross sections for a 12 MeV neutrino. 
The solid lines correspond to the reaction $\nu_e + n \rightarrow e^- + p$ 
and the dashed lines correspond to $\bar \nu_e + p \rightarrow e^+ + n$.  
The black lines are calculations in which mean field effects have been included,
while the red lines are calculations in which the mean field effects have 
been ignored.  The green dotted line corresponds to the available electron phase 
space, arbitrarily scaled.  The assumed background conditions are $T$ = 8 MeV, 
and $n_B = 0.02 \, {\rm fm}^{-3}$.  The electron fraction is 0.027, which corresponds 
to beta equilibrium for the given temperature, density, and the assumed nuclear 
interactions.  The nucleon potential difference is $U_n-U_p=\Delta U=9 \, {\rm MeV}$.
All cross-sections are for the same baryon density and electron fraction (i.e. all 
assume the same $\tilde \mu$ for the neutrons and protons). }
\label{fig:diff_csec}
\end{center}
\end{figure}

Because $\Delta U \gtrsim \epsilon_\nu$ and $T$ for neutrino energies of interest in the decoupling region, it introduces strong asymmetry between the electron neutrino and antineutrino charged current interactions because the $Q$ value for the reaction $\nu_e + n \rightarrow e^- + p$ is increased by $\Delta U=U_n-U_p$ and for $\bar \nu_e + p \rightarrow e^+ + n$ it is reduced by the same amount. Since $\Delta U < \mu_e$, this amount of energy is often not enough to put the final state electron above the Fermi surface. However, it is enough to put the final state electron in a relatively less blocked portion of phase space resulting in an exponential enhancement of the cross-section for $\nu_e$.  This is shown in Figure \ref{fig:diff_csec}, where the differential cross-section integrated over angle for charged current absorption is plotted as a function of the final lepton energy.  The neutrino energy is set to $12$ MeV and the conditions of the medium are
$T$ = 8 MeV, and $n_B = 0.02 \, {\rm fm}^{-3}$ and $Y_e=0.027$.  The peak of the differential cross-section is shifted by about $\Delta U$ up (down) in $\varepsilon_{e^-}$ 
($\varepsilon_{e^+}$) for electron (anti-)neutrino capture.  This shift significantly increases the available phase space for the final state electron in $\nu_e + n 
\rightarrow e^- + p$.  The (arbitrarily scaled) phase space factor $p_e E_e (1-f_e)$ is also plotted and the peak of $1/Vd\sigma/dq_0$ approximately follows this relation.  
As was argued in section \ref{sec:kinematics}, the rate of $\bar \nu_e + p \rightarrow e^+ + n$ should also be approximately proportional to this phase space factor and be
exponentially suppressed.  This is seen in the Figure \ref{fig:diff_csec}.

\begin{figure}
\begin{center}
\leavevmode
\includegraphics[width=\columnwidth]{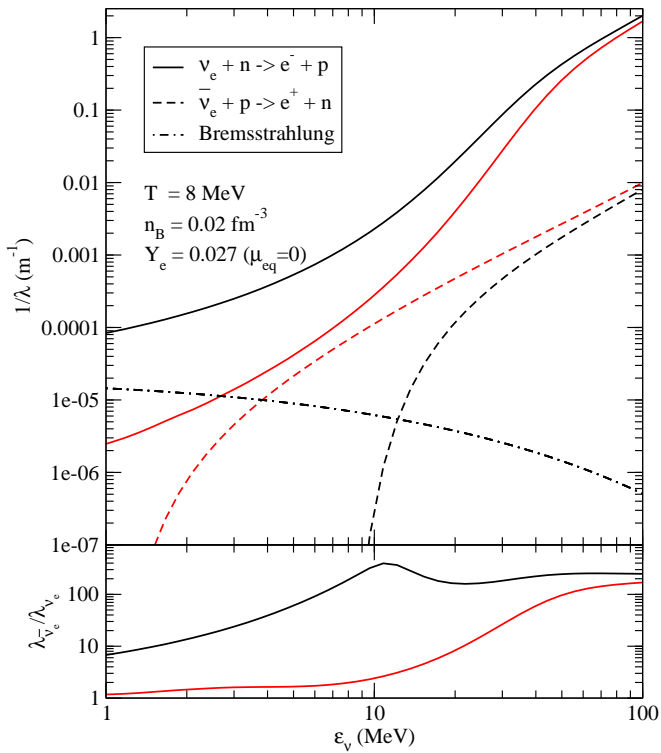}
\caption{
The top panel shows the total absorption inverse mean free path as a function of incoming neutrino 
energy for electron neutrinos (solid lines) and electron antineutrinos (dashed lines).  The 
dot-dashed line shows the effective bremsstrahlung inverse mean free path.  In both panels 
the black lines include mean field effects and the red lines assume a free gas response function.  
The bottom panel shows the ratio of the total electron neutrino capture rate to the total electron
antineutrino capture rate.  Beta-equilibrium has been assumed and the temperature has 
been fixed at 8 MeV.}
\label{fig:mfp}
\end{center}
\end{figure}
In Figure \ref{fig:mfp}, the inverse mean free path ($\lambda^{-1} = \sigma/V$) is shown as a function 
of neutrino energy for the same conditions considered in Figure \ref{fig:diff_csec}.  At low energies 
the electron neutrino mean free path is reduced when mean fields are correctly incorporated, but at 
larger neutrino energies the presence of mean fields becomes less important and the mean free paths with
and without mean fields asymptote to each other.  The electron antineutrino mean free path is reduced relative
to the free gas result and the presence of a threshold at the potential difference $\Delta U$ is evident 
in the mean field calculation.  The effective bremsstrahlung mean free path is also plotted.  This is calculated
assuming the secondary neutrinos are in thermal equilibrium with the background, which is a good approximation
for electron antineutrino destruction.  For electron antineutrinos at low energies, bremsstrahlung dominates the 
capture rate.  Mean field effects push the energy region were bremsstrahlung is dominant to larger
neutrino energies.  This suggests that varying the assumed bremsstrahlung rate will also affect the 
spectrum of the electron antineutrinos.  In the bottom panel, the ratio of the electron antineutrino 
mean free path to 
the electron neutrino mean free path is shown as a function of energy with and without the affect of mean
fields.  The large asymmetry induced between electron neutrino and antineutrino 
charged current interactions when mean fields are properly included is plainly visible.

The formalism of \cite{Reddy98} includes this effect, and was used to calculate the neutrino interaction
rates employed in the models presented in \citep{Roberts12} and in section \ref{sec:PNS_evolve} of this 
work.  However, the formulae in \cite{Bruenn85} and \cite{Burrows99} for charged current rates neglect 
the potential energy difference in the nucleon kinematics.  
In \cite{Burrows99}, a procedure is advocated for including mean fields in 
which the effective chemical potential, $\tilde \mu_i$ of each species is calculated from the given number 
density and temperature by inverting the free Fermi gas relation, then the response is assumed to be the 
free gas response but with the effective chemical potentials in place of the actual chemical potentials. 
This prescription is incorrect because while it accounts for the location of the Fermi surface of the nucleons 
it fails to account for the presence of a potential energy difference between incoming and outgoing nucleon 
states.  This amounts to assuming $\mu \rightarrow \tilde \mu$, so that in Eq.~\ref{eq:response} $\tilde 
q_0 \rightarrow q_0$ and the response becomes the non-interacting response for the given density and electron 
fraction.  When the potential energies of the incoming and outgoing nucleons states are equal, as in symmetric 
matter, or for neutral current reactions  this prescription results in the correct expression, but in asymmetric 
matter and for charged current reactions it is in error. To obtain the correct expression for the mean field   
polarization function from the free gas results of \cite{Burrows99} it is necessary to make {\it both} replacements 
given in Eq. \ref{eq:mf_replacements}.

\subsection{Correlations and Collisional Broadening}
In addition to the mean field energy shift, interactions correlate and scatter nucleons in the medium. 
The excitation of two or more nucleons by processes such as $\nu_e+n+n\rightarrow n+p+e^-$ and $\nu_e+n+p \rightarrow p+p+e^-$ alter the kinematics of the charged current reaction. Typically, these two-particle reactions introduce modest corrections to the single-particle response when the quasi-particle life-time is large. However, they can dominate when: (i) energy-momentum requirements are not fulfilled by the single particle reaction; (ii) final state Pauli blocking requires large energy and momentum transfer; (iii) or both. Such circumstances are encountered in neutron star cooling, where the reaction $n\rightarrow e^-+p+\bar{\nu}_e$ is kinematically forbidden at the Fermi surface under extreme degeneracy unless the proton fractions $x_p \gtrsim 10\%$ \citep{Lattimer91,Pethick92}. Instead, the two-particle reaction $n + n \rightarrow e^- + p +\bar{\nu}_e$, called the modified URCA reaction, is the main source of neutrino production \citep{Friman79}. At temperatures encountered in PNS cooling, energy-momentum restrictions do not forbid the single-particle interactions, but they do strongly frustrate them due to final state blocking.  

The excitation of two particle states in neutral current reactions has been included in a unified approach described in \cite{Lykasov08} and incorporated into the total response function by introducing a finite quasi-particle lifetime $\tau$. This naturally leads to collisional broadening allowing the response to access multi-particle kinematics and alters both the overall shape and magnitude of the response function \citep{Hannestad98,Lykasov08}.  Here, as a first step, we adapt the general structure of the response function from \cite{Lykasov08} to show that two-particle excitations play an important role in the charged  current process.  We include a finite $\tau$ through the following ansatz for the imaginary part of the polarization function 
\bea
\text{Im}~\Pi_\Gamma &=& -2 \pi \int\frac{d^3p}{(2\pi)^3} \frac{T~z~[f_4(\epsilon_{p+q})-f_2(\epsilon_p)]}{\Delta\epsilon_{p+q}+\hat{\mu}-\Delta U} \text{L}(\Gamma) ~~~\\
\text{L}(\Gamma)&=&\frac{1}{\pi}\frac{\Gamma}{(\tilde{q}_0-\Delta\epsilon_{p+q})^2+\Gamma^2}\,,
\label{eq:lps_charged}
\eea 
which is obtained by replacing the energy delta-function in the Fermi gas  particle-hole polarization function (see Eq.~\ref{eq:response} and Eq.~\ref{eq:pi_fg}) by a Lorentzian with a width $\Gamma=1/\tau$.  Here, as before $z=(q_0+\hat{\mu})/T$ and $\Delta\epsilon_{p+q}=\epsilon_{p+q}-\epsilon_p$.The Lorentzian form is obtained in the relaxation time approximation discussed in  \cite{Lykasov08}, and is valid when $|q_0| \tau \gg 1$.  The quasiparticle lifetime $\tau$ is a function of the quasi-particle momentum, $q$, $q_0$ and the ambient conditions. Its magnitude and functional form at long-wavelength is constrained by conservation laws. For the vector-response, $\tau\rightarrow \infty $ in the limit $q\rightarrow 0$ due to vector current conservation.  However, because spin is not conserved by strong tensor and spin-orbit interactions, the nucleon spin fluctuates even at $ q\rightarrow 0$ and the associated spin relaxation time $\tau_\sigma$ is finite \citep{Hannestad98}. Since the spin response dominates the charged current reaction, in what follows we shall use Eq.~\ref{eq:lps_charged} only to modify the spin part of the charged current response. We, however, note that the multi-particle response in the vector channel warrants further study since the typical momentum transfer $q\simeq \mu_e$ is not negligible.   

For the spin  relaxation time $\tau_\sigma$ we use results calculated in Ref.~\cite{Bacca11} which indicate that it decreases rapidly with both density and temperature. The typical range of values of $1/\tau_\sigma$ obtained from \cite{Bacca11} but including a 50\% variation over their quoted values is shown in Figure \ref{fig:potentials} for conditions in the neutrino sphere region.  Using these values as a guide we study the effects of collisional broadening on the $\nu_e$ and $\bar{\nu}_e$ cross-sections.  
\begin{figure}
\begin{center}
\includegraphics[width=\columnwidth]{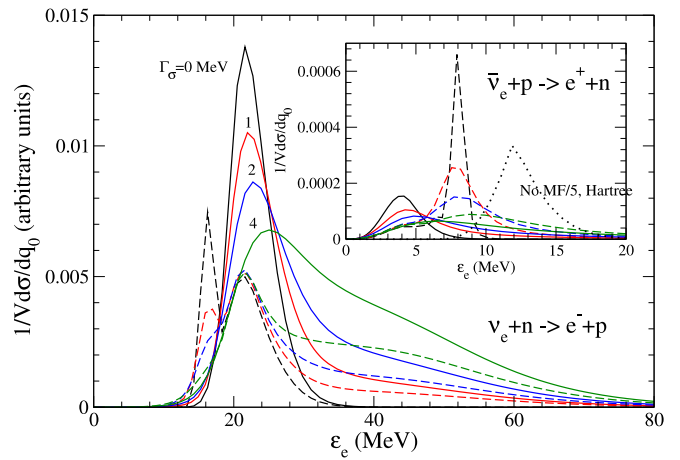}
\caption{The axial portion of the $\nu_e$ (main panel) and $\bar{\nu}_e$ (inset) absorption cross-section 
including collisional broadening.  This shifts a significant fraction of the response to larger $E_e$
where there is larger lepton phase space available. The ambient conditions and neutrino energy are the same as those in figure \ref{fig:diff_csec}.  The dashed lines show the RPA response, including both mean fields and collisional 
broadening.  The dotted line in the inset panel shows the free gas response.}
\label{fig:dsigma_tau}
\end{center}
\end{figure} 
The differential cross-section for the axial portion of the process $\nu_e+n\rightarrow p+e^-$ is shown in
Figure \ref{fig:dsigma_tau} for $T$ = 8 MeV, $n_B = 0.02 \, {\rm fm}^{-3}$, and $Y_e=0.027$.  The initial neutrino energy is $E_{\nu_e}=12$ MeV.  As before the differential cross-section is plotted as function of the outgoing electron energy. The result with $\Gamma \rightarrow 0$ recovers the single-particle response with the mean field energy shift included. Representative values of $\Gamma=1,2,4$ MeV are chosen to approximately reflect the findings of \cite{Bacca11} for these ambient conditions.  The collisional broadening seen in Figure \ref{fig:dsigma_tau} is quite significant. It increases the the axial portion of the cross-section by approximately $20\%,44\%$ and $80\%$, for $\Gamma=1,2,4$ MeV, respectively.  Together, the mean field energy shift and collisional broadening push strength to regions where electron final state blocking is smaller resulting in an overall increase in the electron neutrino absorption rate.       

While mean field effects reduce the $\bar{\nu}_e$ cross-section, collisional broadening will tend to increase it by accessing kinematics where $-q_0$ is larger. This is shown in the inset of Figure \ref{fig:dsigma_tau} where the $\bar{\nu}_e$ cross-section for the same ambient conditions and for $E_{\bar{\nu}_e}=12$ MeV is plotted as a function of the positron energy $E_e^+=E_\nu-q_0$ . The units are arbitrary and the plots only serve to illustrate the relative effect of multi-pair excitations. We choose the same values of $\Gamma$ as for the $\nu_e$ case. Here broadening due to  multi-pair excitations has a more significant effect than for $\nu_e$ absorption. However, despite this enhancement, the response that includes the mean field energy shift and collisional broadening is still much smaller than the free gas response. 

In addition to multi-pair processes, weak charge screening in the medium can also affect the charged current response. 
Screening due to correlations has been investigated in the Random Phase Approximation (RPA), where specific long-range correlations are included by summing single-pair ``bubble'' or particle-hole diagrams.  Additionally, this approach ensures consistency between the response functions and the underlying equation of state in the long wavelength limit.  For charged currents, calculations reported in \cite{Burrows99} and \cite{Reddy99} indicate that the suppression is density and temperature dependent. It can be as large as a factor of $2$ at supra nuclear density, but at densities of relevance to the neutrino sphere where $\rho \lesssim 10^{13}$ g/cm$^3$ the corrections are $\approx 20\%$. More importantly,  the suppression found in \cite{Burrows99} and \cite{Reddy99} for the charged current rate is a weak function of reaction kinematics and can viewed as a overall shift of the response in Fig.~\ref{fig:diff_csec}, aside from regions were significant strength is shifted to collective modes. The energy and momentum restrictions discussed previously apply also to the RPA response, and the mean field energy shift is important to include in the calculation of the particle-hole diagrams. They were included in \cite{Reddy99} but omitted in  \cite{Burrows99}.  

To include correlations between particle-hole (p-h) excitations due to residual interactions in the spin-isospin channel using the RPA, we employ a constant interaction (independent of momentum and density) in the spin-independent and spin-dependent particle-hole channels given by  $V^\text{ph}_{\tau}=2$ fm$^{-2 }$ and $V^\text{ph}_{\sigma \tau}=1.1$ fm$^{-2 }$, respectively.  The residual interaction in spin-independent channel is consistent with underlying equation of state. The residual interaction in spin-dependent particle-hole channel is retrieved from analysis of the Gamow-Teller transition in finite nuclei \citep{Bertsch81}.  The RPA response functions for this simple form of the p-h interaction are then given by
\bea
S_\tau(q_0,q)&=& \frac{2}{1 - e^{-z}}~\text{Im}\left[\frac{\Pi_\text{MF}}{1-V^\text{ph}_{\tau}\Pi_\text{MF}}\right]\\
S_{\sigma \tau}(q_0,q)&=& \frac{2}{1 - e^{-z}}~\text{Im}~\left[\frac{\Pi_\Gamma}{1-V^\text{ph}_{\sigma \tau}\Pi_\Gamma}\right]\,,
\eea
and the real and imaginary parts of the polarization functions satisfy the Kramers-Kronig relation, 
\be 
\text{Re}~\Pi(q_0,q)=-\frac{{\cal P}}{\pi}\int~d\omega~ \frac{\text{Im}~\Pi(\omega,q)}{\omega-q_0} \,. 
\ee
RPA correlations also act to redistribute the strength of the response.  The RPA response is shown in Figure \ref{fig:dsigma_tau}.  The Gamow-Teller resonance is clearly visible in the curves that do not include large 
amounts of collisional broadening.  The extent to which this affects the inverse mean free paths can be gauged from the results presented in Table \ref{tab:sum}.  

\begin{table*}[htbp]
\centering
\caption{$1/\lambda$ in m$^{-1}$ for matter in beta-equilibrium at $T= 8$ MeV and various densities and $E_{\nu_e}=E_{\bar{\nu}_e}=12$ MeV.  The entries in the table follow the notation $\scit{a}{b} = \sci{a}{b}$.  In the last two 
columns, $\Gamma$ is considered to density dependent and the values used are taken from Figure \ref{fig:potentials}.} 
\begin{tabular}{@{} lcccccc @{}} 
\hline
	Density (fm$^{-3}$) 	& $1/\lambda $ (m$^{-1}$):		&  no MF 		&	MF  ($\Gamma=0$) & 	RPA ($\Gamma=0$) &	MF ($\Gamma>0$)& 	RPA ($\Gamma>0$) \\ \hline
	$n_B=0.020$ 			& $1/\lambda_{\nu_e}:$			& $\scit{5.9}{-4}$ & $\scit{5.2}{-3}$ & $\scit{2.1}{-3}$ & $\scit{7.5}{-3}$ & $\scit{3.9}{-3}$ \\
						& $1/\lambda_{\bar{\nu}_e}:$		& $\scit{3.5}{-4}$ & $\scit{2.7}{-5}$ & $\scit{6.5}{-5}$ & $\scit{4.5}{-5}$ & $\scit{6.0}{-5}$ \\
\hline
	$n_B=0.006$ 			& $1/\lambda_{\nu_e}:$    		& $\scit{7.7}{-4}$ & $\scit{1.6}{-3}$ & $\scit{1.2}{-3}$ & $\scit{1.8}{-3}$ & 	$\scit{1.3}{-3}$ \\
						& $1/\lambda_{\bar{\nu}_e}:$     & $\scit{2.4}{-4}$ & $\scit{1.4}{-4}$ & $\scit{2.0}{-4}$ & $\scit{1.4}{-4}$ & $\scit{1.5}{-4}$ \\
\hline
	$n_B=0.002$			& $1/\lambda_{\nu_e}:$ 			& $\scit{5.3}{-4}$ & $\scit{6.5}{-4}$ & $\scit{5.9}{-4}$ & $\scit{6.8}{-4}$ & 	$\scit{6.1}{-4}$ \\     
						& $1/\lambda_{\bar{\nu}_e}:$     & $\scit{1.5}{-4}$ & $\scit{1.3}{-4}$ & $\scit{1.4}{-4}$ & $\scit{1.3}{-4}$ & 	$\scit{1.3}{-4}$ \\     
\hline
\end{tabular}
\label{tab:sum}
\end{table*}

While collisional  broadening tends to increase both $\nu_e$ and $\bar{\nu}_e$ cross-sections, RPA correlations decrease the $\nu_e$ cross-section and enhance the cross-section for $\bar{\nu}_e$. Given the simplicity of our model for the p-h interaction, these results only serve to capture the qualitative aspects of the role of correlations. They nonetheless demonstrate that changes expected are small compared to corrections arising due to a proper treatment of mean field effects in the reaction kinematics. Hence, in the following discussion of PNS evolution and neutrino spectra, we set aside these effects due to RPA correlations and collisional broadening, and calculate the neutrino interactions only including the mean field energy shifts calculated as described in \cite{Reddy98}.    

\section{Proto-Neutron Star Evolution}
\label{sec:PNS_evolve}
To illustrate the effect of the correct inclusion of mean field effects
in charged current interaction rates, as well as the importance of the 
nuclear symmetry energy, five PNS cooling models are described here.  The 
models have been evolved using the multi-group, multi-flavor, general 
relativistic variable Eddington factor code described in \citep{Roberts12b}
which follows the contraction and neutrino losses of a PNS over the first 
$\sim 45$ seconds of its life.  These start from the same post core 
bounce model considered in \citep{Roberts12b} and follow densities down 
to about $10^9 \, {\rm g \, cm}^{-3}$.  Therefore, they do not simulate the NDW 
itself but they do encompass the full neutrino decoupling region. 
\begin{figure}
\begin{center}
\leavevmode
\includegraphics[width=\columnwidth]{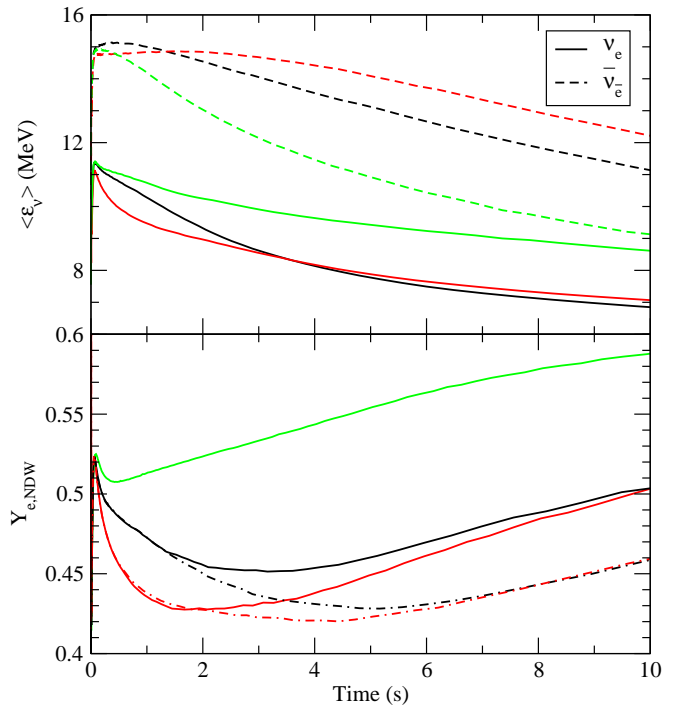}
\caption{
{\it Top panel}: First energy moment of the outgoing electron neutrino
and antineutrino as a function of time in three PNS cooling simulations.
The solid lines are the average energies of the electron neutrinos and 
the dashed lines are for electron antineutrinos.  The black lines correspond
to a model which employed the GM3 equation of state, the red lines to 
a model which employed the IU-FSU equation of state, and the green
lines to a model which ignored mean field effects on the neutrino opacities
(but used the GM3 equation of state).  
{\it Bottom panel}: Predicted neutrino driven wind electron fraction
as a function of time for the three models shown in the top panel (solid lines), 
as well as two models with the bremsstrahlung rate reduced by a factor of 
four (dot-dashed lines).  The colors are the same as in the top panel.}
\label{fig:Average Energy Compare}
\end{center}
\end{figure}

One model was run using neutrino interaction rates that ignore the presence 
of mean fields, but are appropriate to the local nucleon number densities 
(i.e. the re-normalized chemical potentials, $\tilde \mu_i$, were used but 
we set $\Delta U = 0$).  The equation of state used was GM3.  This model was 
briefly presented in \citep{Roberts12b}.  Another model was calculated that 
incorporated mean field effects in the neutrino interaction rates and used 
the GM3 equation of state.  A third model was run using the IU-FSU equation of state 
and including mean field effects but with everything else the same as the 
GM3 model.  Additionally, two similar models were run with the bremsstrahlung 
rates of \citep{Hannestad98} reduced by a factor of 4 as suggested by 
\citep{Hanhart01}.  The neutrino interaction rates in all five models were calculated
using the relativistic polarization tensors given in \citep{Reddy98} with the weak
magnetism corrections given in \citep{Horowitz03}.

In the top panel of Figure \ref{fig:Average Energy Compare}, the average
electron neutrino and antineutrino energies are shown as a function of time 
for the three models with the standard bremsstrahlung rates.  As was described 
in \citep{Roberts12b}, including mean 
field effects in the charged current interaction rates significantly reduces 
the average electron neutrino energies because the decreased mean free paths
(relative to the free gas case) cause the electron neutrinos to decouple at 
a larger radius in the PNS and therefore at a lower temperature.  Conversely,
for the electron antineutrinos the mean free path is increased, they decouple
at a smaller radius and higher temperature, and therefore their average energies 
are larger.  Mean field effects serve to shift the average neutrino energies 
by around 25\% at later times.  The antineutrino energies are also slightly 
larger than the values reported in \citep{Roberts12b} because of the reduced 
bremsstrahlung rate.

To illustrate the properties of the region where neutrino decoupling occurs, 
a snapshot of the decoupling region as a function of neutrino energy is shown 
in Figure \ref{fig:decoupling}.  In this work, the ``decoupling region'' is 
defined as the region where the Eddington factor $f_1=F_g/N_g$ obeys the 
condition $0.1<f_1<0.5$.  Here, $F_g$ is the neutrino number flux in energy 
group $g$ divided by the speed of light and $N_g$ is the neutrino number density
in energy group $g$ (see \citep{Roberts12b}).  This approximately defines the 
region over which neutrinos transition from being diffusive to free-streaming.
Higher energy electron neutrinos decouple at a larger radius and therefore a 
lower density and temperature.  At these radii, $\Delta U$ is smaller than the 
temperature and the inclusion of mean fields in the interaction rates should 
not significantly change the high energy electron neutrino mean free paths.  At 
lower neutrino energies, $\Delta U$ is significantly larger than the temperature
in the decoupling region and the presence of mean fields strongly affects the 
opacity.  As time progresses, the average neutrino energies become lower and 
decoupling occurs in conditions at which mean field effects become increasingly
important.  Decoupling also occurs at a higher density for lower energy neutrinos, 
where both multi-particle processes and RPA corrections can potentially become 
important.

\begin{figure}
\begin{center}
\includegraphics[width=\columnwidth]{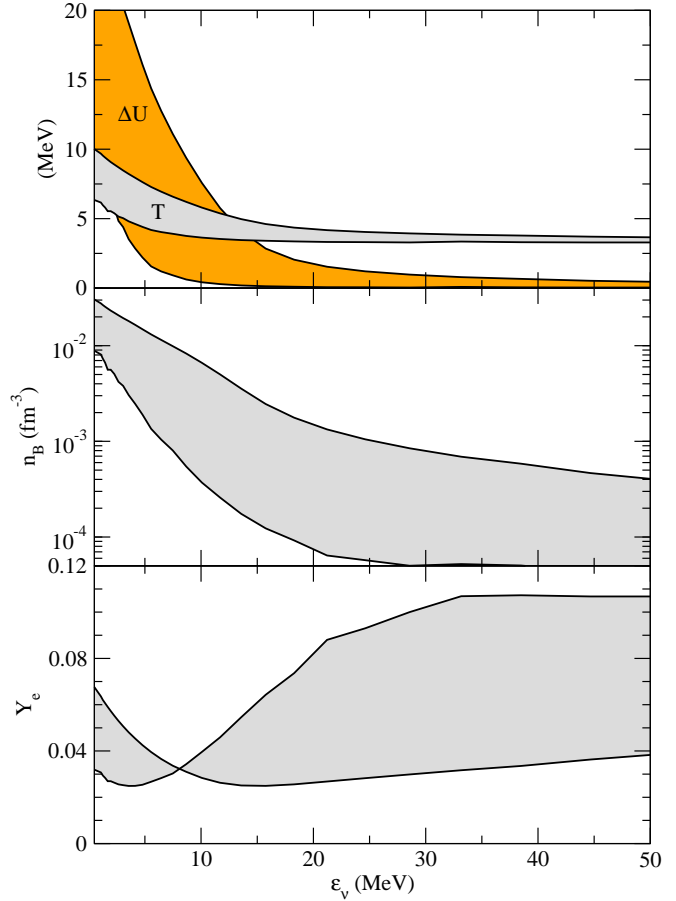}
\caption{The thermodynamic conditions and nucleon potential difference 
characterizing the region where electron neutrinos decouple as a function of 
neutrino energy.  These values are taken from the PNS model which employed the 
IU-FSU equation of state at 3.3 seconds after core-bounce.  At this point the 
average electron neutrino energy is 8.3 MeV.}
\label{fig:decoupling}
\end{center}
\end{figure} 

Additionally, there are significant differences between the two models which 
include mean field effects but use different equations of state.  As was 
described above, the GM3 equation of state has a smaller symmetry energy than 
the IU-FSU equation of state at sub-nuclear densities and therefore has a 
smaller $\Delta U$ in the neutrino decoupling region.  This suggests that GM3 
should have slightly larger electron neutrino average energies and slightly 
lower average electron antineutrino energies.  The results of self-consistent
PNS simulations are somewhat more complicated than this simple picture, mainly
because the equilibrium electron fraction near the neutrino sphere also depends
on the nuclear symmetry energy which affects the charged current rates (see 
Figure \ref{fig:potentials}).  Still, there is a larger difference between the
average electron neutrino and antineutrino energies throughout the simulation 
(relative to GM3) when the IU-FSU equation of state is used, as expected.  

The moments of the escaping neutrino distribution along with the electron 
neutrino number luminosities can be used to calculate an approximate NDW 
electron fraction \citep{Qian96} 
\be
\label{eq:Ye_NDW}
Y_{e,{\rm NDW}} \approx \left [1 + \frac{\dot N_{\bar \nu_e}
\langle \sigma(\epsilon)_{p,\bar\nu_e}\rangle}{\dot N_{\nu_e}\langle 
\sigma(\epsilon)_{n,\nu_e}\rangle}  \right]^{-1},
\ee 
where $\dot N$ are the neutrino number luminosities and $\langle \sigma(\epsilon)\rangle$ 
are the energy averaged charged current cross-sections in the wind region.  
The approximate NDW electron fraction as a function of time for the five models is
shown in the bottom panel of Figure \ref{fig:Average Energy Compare}.  The low 
density charged current cross-sections from \citep{Burrows06} were used.  First, 
it is clear from this plot that mean field effects significantly decrease the 
electron fraction in the wind.  This is mainly due to the increased difference 
between the electron neutrino and antineutrino average energies caused by the 
effective $Q$ value induced by the mean field potentials. Second, increasing the 
sub-nuclear density symmetry energy decreases the electron fraction in the wind. 
This in turn implies that nucleosynthesis in the NDW may depend on the nuclear 
symmetry energy because it is sensitive to electron fraction in the wind 
\citep[e.g.][]{Hoffman97}.  Still, this effect is not particularly strong because 
the increase in the electron neutrino cross section for increased $\Delta U$ is 
partially mitigated by the larger equilibrium electron fraction predicted for models 
with a larger nuclear symmetry energy.

\section{Conclusions} 
In this work, we have discussed the physics of charged current neutrino 
interactions in interacting nuclear matter at densities and temperatures 
characteristic of the neutrino decoupling region in PNS cooling.  Additionally,
models of PNS cooling have been run to assess the importance of changes 
in the charged current rates to the properties of the emitted neutrinos.
Our main findings are:

\begin{itemize}
\item The mean-field shift of the nucleon energies  alters the kinematics of 
the charged current reactions. Under neutron-rich conditions it increases the 
$Q$-value for $\nu_e$ absorption and decreases it for $\bar{\nu}_e$. 
Due to final state blocking (electron blocking for electron neutrino capture
and neutron blocking for electron antineutrino capture), the increase in the 
$Q$ value leads to an exponential ($\exp{(\Delta U/T)}$) increase in the 
$\nu_e$ cross-section absorption and reduces the $\bar{\nu}_e$ absorption 
cross-section by $\exp{(-\Delta U/T)}$. 

\item The formulae for the charged rates developed in \cite{Burrows99} and
\cite{Bruenn85} neglect these effects and the prescription for incorporating 
mean field energy shifts outlined in \cite{Burrows99} is inconsistent. 

\item The nuclear symmetry energy at sub-nuclear density plays a crucial role in 
determining the magnitude of the difference between the mean field neutron and 
proton potential energies, and through its effect on the $Q$-values increases 
the difference between the mean free paths of $\nu_e$ and $\bar{\nu}_e$.  This 
sensitivity to the symmetry energy is potentially exciting since supernova neutrino 
detection and nucleosynthetic yields may be able to provide useful constraints.    

\item Our {\it preliminary} work indicates that multi-pair excitations favor 
kinematics where final state electron blocking is small because the 
energy/momentum constraints present when only single particle-hole (p-h) excitations 
are considered are relaxed.  This is analogous to the importance of the modified 
URCA process in neutron star cooling.  In contrast to mean field effects, multi-pair
excitations decrease the mean free paths of both electron neutrinos and electron
antineutrinos.  

\item Nuclear correlation effects treated in the RPA decrease the 
$\nu_e$ cross-section and enhance the cross-section for $\bar{\nu_e}$. However, 
preliminary calculations using residual interactions consistent with equation of state or derived 
from Gamow-Teller transitions of finite nuclei suggest that the changes are much smaller 
than the proper inclusion of mean field effects in the reaction kinematics.  Although it 
is difficult to determine from the limited and approximate calculations performed for 
this work, it seems most likely that multi-pair excitations and RPA corrections will bring 
the average electron neutrino and antineutrino energies somewhat closer to one another 
(relative to the case were only mean fields are included).

\item As was shown in \citep{Roberts12b}, the changes to the charged current 
mean free paths induced by the correct inclusion of mean fields decreases
the average energy of the electron neutrinos and increases the average 
energy of the anti-electron neutrinos emitted during PNS cooling.  The difference 
is relatively large, it significantly alters the predicted electron fraction in the 
NDW, and may have observable effects.  This result has recently been independently 
confirmed by \cite{Martinez-Pinedo12}.

\item We have also directly shown that increasing the value of the nuclear 
symmetry energy at sub-nuclear densities decreases the electron fraction in
the neutrino driven wind.  Therefore, NDW nucleosynthesis may put some constraint
on the poorly known density dependence of the nuclear symmetry energy, or
vice versa.  This potential astrophysical constraint is in addition to those discussed 
in \cite{Lattimer12}.  We emphasize that it may be hard to disentangle this 
from the effects of multi-particle excitations, both on the charged current reactions 
themselves and on the (related) bremsstrahlung rate.  This effect is also partially 
compensated by the symmetry energy dependence of the beta-equilibrium electron 
fraction.

\item The reduced mean free path of $\nu_e$ is also likely to affect the 
de-leptonization time of the proto-neutron star and may account for differences 
in time-scales observed in simulations performed using equations of state with 
different symmetry energies.
\end{itemize}
Our work also shows that  multi-particle excitations and correlations can alter the charged current response by as much as 
as factor of two at densities realized in the neutrino decoupling region. However, our simple treatment has large uncertainty and 
warrants further study before we can make reliable predictions for the difference between $\nu_e$ and $\bar{\nu}_e$ spectra. Since this difference affects nucleosynthesis, collective neutrino oscillations, and is potentially observable from the high statistics expected for a galactic supernova neutrino burst, our study here identifies that there is still
much work to pursue both with respect to the charged current reactions and equation of state of neutron-rich matter in the neutrino decoupling region.  

\begin{acknowledgments}
We gratefully acknowledge George Bertsch, Vincenzo Cirigliano, and Stan Woosley
for useful discussions concerning this work.  We also thank Georg Raffelt for stimulating conversations about neutrino rates in current supernova simulations.  L. R. acknowledges support 
from the University of California Office of the President (09-IR-07-117968-WOOS).  
His research has also been supported at UCSC by the National Science Foundation (AST-0909129). 
The work of S.R. was supported by the DOE grant \#DE-FG02-00ER41132 and by the 
Topical Collaboration to study {\it Neutrinos and nucleosynthesis in hot dense matter}. 
\end{acknowledgments}

\hspace{60mm}

\end{document}